\documentstyle[prb,twocolumn,aps,epsf]{revtex}

\begin{document} 
\draft
\preprint{cond-mat}
\twocolumn[\hsize\textwidth\columnwidth\hsize\csname @twocolumnfalse\endcsname
\title{Persistent Currents in Quantum Chaotic Systems}
\author{Shiro Kawabata}
\address{
Physical Science Division, Electrotechnical Laboratory, 1-1-4 Umezono, Tsukuba, 
Ibaraki 305-8568, Japan
}
%
\date{\today}
\maketitle
\begin{abstract} 

The persistent current of ballistic chaotic billiards is considered with the help of the Gutzwiller trace formula.
We derive the semiclassical formula of a typical persistent current $I^{typ}$ for a single billiard and an average persistent current $\left< I \right>$ for an ensemble of billiards at finite temperature.
These formulas are used to show that the persistent current for chaotic billiards is much smaller than that for integrable ones. The persistent currents in the ballistic regime therefore become an experimental tool to search for the quantum signature of classical chaotic and regular dynamics.

\end{abstract}
\vskip2pc]
\narrowtext
A charged particle moving in the presence of a vector potential acquires the Aharonov-Bohm (AB) phase, leading to an interesting interference effect in a non-simply connected geometry.
One of the most evident manifestations of this AB effect is the persistent current in a mesoscopic ring threaded by magnetic flux.~\cite{rf:PC}
Efforts in this direction have been mostly concerned with the influence of 
disorder~\cite{rf:disorder1,rf:disorder2,rf:disorder3,rf:disorder4,rf:disorder5,rf:disorder6} and electron-electron interaction~\cite{rf:e-e1,rf:e-e2,rf:e-e3} on the persistent current.
The quantum-mechanical magnetic properties of ballistic systems depends on the underlying classical dynamics.
Therefore, the persistent current in these systems, i.e., billiards,  is one of the most interesting models for studying quantum chaos.~\cite{rf:QCmeso}

Recently von Oppen and Riedel~\cite{rf:Oppen} showed that the typical semiclassical persistent current at zero temperature for $single$ integrable billiards, $I^{typ} \sim \left( k_F \ell \right)^{3/2}$, is larger than that for chaotic billiards, $I^{typ} \sim  k_F \ell$, by using dimensional analysis ($k_F$ is the Fermi wave number, and $\ell$ is the typical dimension of the billiard).
Based on these results,  Richter, Ullmo, and Jalabert~\cite{rf:RUJ1} calculated the $average$ persistent current $\left< I \right>$ for an ensemble made of integrable billiards with a dispersion in size or in the electron filling by using the Berry-Tabor formula,~\cite{rf:BerryTabor1,rf:BerryTabor2} and showed that it scales as $\left< I \right> \sim  k_F \ell $. 
Until now, however, explicit semiclassical formulas for persistent current in single chaotic billiards or an  ensemble of chaotic billiards have not been derived.

In this paper, we study  the persistent current for quantum chaotic systems by employing  the Gutzwiller trace formula.~\cite{rf:Gutzwiller}
We calculate  $I$ at a finite temperature, and derive the formulas for both single billiards, $I^{typ}$, and an ensemble of billiards, $\left< I \right>$.
These formulas are used to show that $I^{typ} \sim  k_F \ell $ and $\left< I \right> \sim \left( k_F \ell \right)^0$.
Furthermore, it turns out that the amplitude of $I$ for quantum chaotic systems is much smaller than that for quantum integrable systems.
This means that one-parameter families of periodic orbits existing in integrable systems interfere constructively, and give a total contribution much larger than that of the unstable isolated periodic orbits typical of chaotic systems.

The persistent current, induced by the magnetic flux $\phi$ in the AB geometry (see Fig.~1), can be obtained from the Helmholtz free energy $F(T,\phi,N)$ as
\begin{eqnarray}
    I =  -c \left( \frac{ \partial F }{\partial \phi} \right)_{T,N}
 .
  \label{eqn:e1}
\end{eqnarray}
The Helmholtz free energy $F(T,\phi,N)$ for a fixed number $N$ of electrons and the grand canonical thermodynamic potential $\Omega(T,\phi,\mu)$ are related by means of a Legendre transform
\begin{equation}
    F(T,\phi,N) = \mu N + \Omega(T,\phi,\mu)
    ,
  \label{eqn:e2}
\end{equation}
where $\mu$ is the chemical potential of a particle reservoir.
$\Omega$  is given by
\begin{equation}
    \Omega(T,\phi,N) =  -\frac{1}{\beta} 
      									\int dE \thinspace d(E,\phi) 
										\ln \left( 1 + \exp \left[ \beta (\mu-E) \right] \right)
 ,
  \label{eqn:e4}
\end{equation}
where $d(E,\phi)$ is one-particle density of states and $\beta=1/k_B T$.
Within  the semiclassical approximation, $d(E,\phi)$ can be  decomposed into mean (Weyl) and oscillating parts, i.e.,
%
%
%
\begin{figure}[b]

\hspace{2.5cm}
\vspace{-1.0cm}
\epsfxsize=4.0cm
\epsfbox{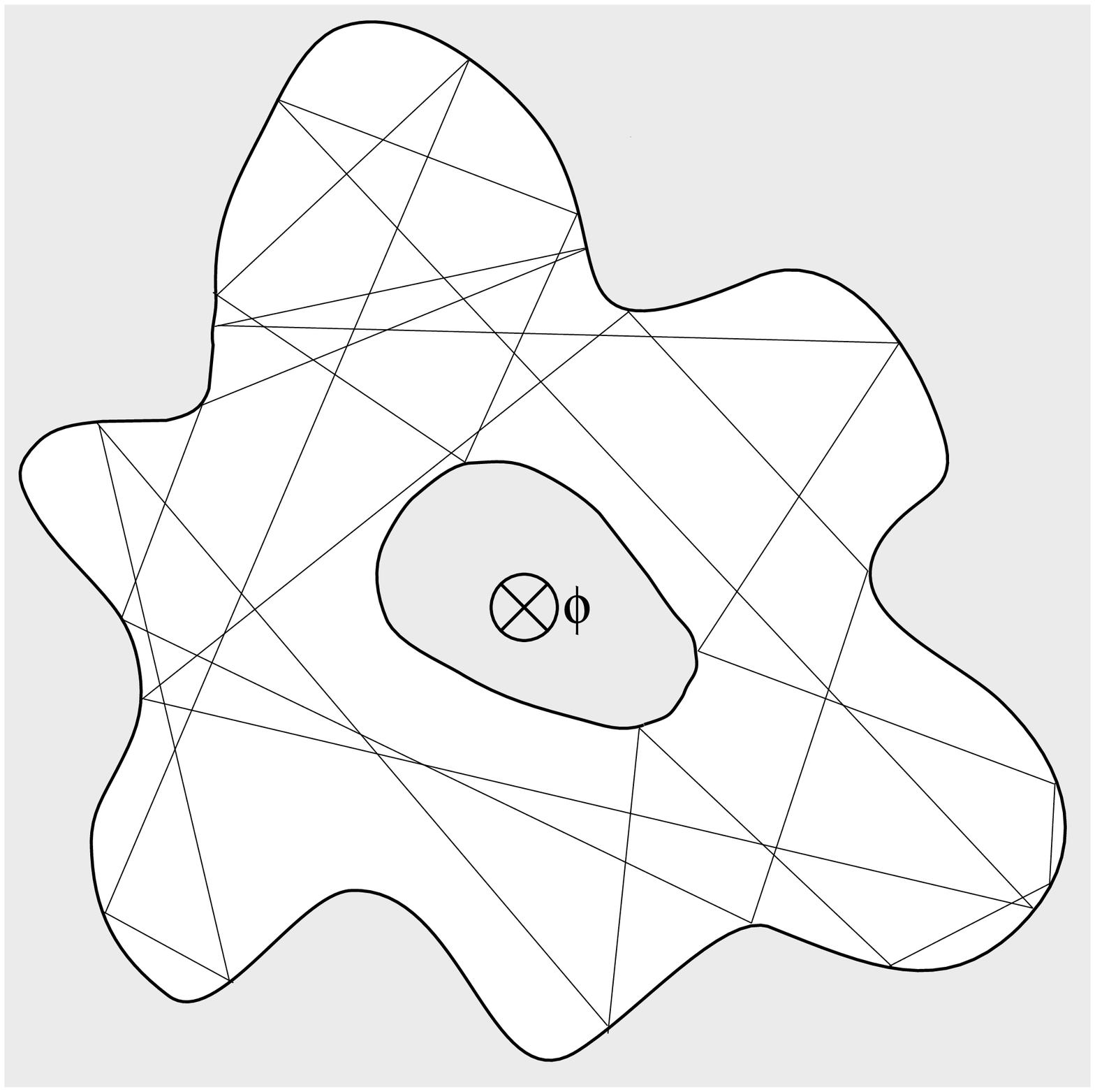}

\caption{A schematic illustration of a chaotic billiard with a rough boundary
threaded by AB flux $\phi$.
An example of an unstable periodic orbit is shown.
}
\end{figure}
%
%
%
%
\begin{equation}
  d(E) =d^0(E) + d^{osc}(E,\phi) 
   .
  \label{eqn:e5}
\end{equation} 
The Weyl term depends on the classical phase volume and is independent of $\phi$.
(The higher $\hbar$ corrections to $d^0$ give the standard bulk Landau magnetism.~\cite{rf:RUJ1})
$d^{osc}(E)$ can be written as a sum over classical periodic orbits $s$,
\begin{eqnarray}
	d^{osc}  (E,\phi) &=& \frac{1}{\pi \hbar} 
	                             \sum_{s} \frac{\tau_{s}}{\left| \det (M_{s} - I )\right|^{1/2}}
								 \nonumber\\
	                             &\times& \cos \left( 
                                            \frac{S_s}{\hbar} + 2 \pi \omega_s \frac{\phi}{\phi_0} 
                                                                         - \sigma_s \frac{\pi}{2} 
                                        \right)
   ,
  \label{eqn:e6}
\end{eqnarray}
where $\tau_{s}$, $M_{s}$, $S_s$, $\omega_s$, and $\sigma_s$ are the period, the monodromy matrix, the action integral, the winding number and Maslov index, respectively.
Following Richter, Ullmo, and Jalabert,~\cite{rf:RUJ1} we define a mean chemical potential $\bar{\mu}$ by the condition of accommodating $\mbox{\boldmath $N$}$ electrons to the mean  number of states,
\begin{equation}
 \mbox{\boldmath $N$}=N(\mu)=\bar{N}(\bar{\mu})  
.
 \label{eqn:e7}
\end{equation}
Here
\begin{equation}
 N(\mu)=\int_{0}^{\infty} dE \thinspace d(E) f(E-\mu)
,
 \label{eqn:e8}
\end{equation}
where $f(E-\mu)$ is the Fermi-Dirac distribution function.
$\bar{N}(\bar{\mu})$ is obtained in eq~(\ref{eqn:e8}) by replacing $d(E)$ and $\mu$ by $d^{0}(E)$ and $\bar{\mu}$.
The Helmholtz free energy in terms of grand canonical quantities is obtained by expanding eq.~(\ref{eqn:e2}) to second order in $\mu-\bar{\mu}$ as
\begin{equation}
 F(N) \approx F^0 + \Delta F^{(1)} + \Delta F^{(2)}
,
 \label{eqn:e9}
\end{equation}
with
\begin{eqnarray}
  F^0 &=& \bar{\mu} \mbox{\boldmath $N$} + \bar{\Omega}(\bar{\mu})
  ,
  \label{eqn:e10}\\
  \Delta F^{(1)} &=& \sum_{s} R_{T}(L_s) \left( \frac{\hbar}{\tau_s} \right)^2 
                               d^{osc}(\bar{\mu},\phi)
  ,
  \label{eqn:e11}\\ 
  \Delta F^{(2)} &=& \frac{1}{2 \bar{d} (\bar{\mu})} 
                              \left[ 
                                     \sum_{s} \frac{-R_T(L_s)}
                                                             {\pi \left| \det (M_{s} - I ) \right|^{1/2} }
							\right.
                            \nonumber\\
	                             &\times& 
							\left.								 
							\sin \left( 
                                            \frac{S_s}{\hbar} + 2 \pi \omega_s \frac{\phi}{\phi_0} 
                                                                         - \sigma_s \frac{\pi}{2} 
                                        \right)
                              \right]^2
,
  \label{eqn:e12}
\end{eqnarray}
where $L_s= v_F \tau_s$ is the length of trajectory $s$, and $\bar{d} (\bar{\mu})=mA/2\pi\hbar^2$ ($m$ is the mass of electron and $A$ is the area of the billiard).
In eqs.~(\ref{eqn:e11}) and (\ref{eqn:e12}),
\begin{equation}
   R_T(L_{s}) = \frac{L_{s}/L_{c} } {\sinh \left( L_{s}/L_{c}  \right)  } 
    ,
 \label{eqn:e13}
\end{equation}
with $L_c= \hbar v_F \beta /\pi$, yields an exponential suppression for the contribution of trajectories with $L>L_c$.

The leading-order contribution to $F$ is given by $F^0 + \Delta F^{(1)}$, and yields the persistent current of {\it single} chaotic billiard calculated in the grand canonical ensemble as
\begin{eqnarray}
	I^{(1)}  (\phi) &=& \frac{e}{\pi} v_F
	                             \sum_{s} 
	                             \frac{ R_t(L_{s}) \omega_s}{L_s \left| \det (M_{s} - I )\right|^{1/2}} \nonumber\\
	                          &\times&   \sin \left( k_F L_s  + 2 \pi \omega_s \frac{\phi}{\phi_0} 
	                                                             - \sigma_s\frac{\pi}{2}  \right)
,
 \label{eqn:e14}
\end{eqnarray}
where $v_F=\hbar k_F /m$ is the Fermi velocity.
Since the system is time reversal invariant at zero magnetic flux, each periodic orbit is associated with a time-reversed partner having exactly the same characteristics except for an opposite winding number.
Grouping the time-reversal trajectories $t$ ($\omega_t>0$) in eq.~(\ref{eqn:e14}), we have
\begin{eqnarray}
	I^{(1)}  (\phi) &=& \frac{2 e}{\pi} v_F
	                             \sum_{t} 
	                             \frac{ R_t(L_{t}) \omega_t}{L_t \left| \det (M_{t} - I )\right|^{1/2}} \nonumber\\
	                         &\times&    \cos \left( k_F L_t  - \sigma_t\frac{\pi}{2}  \right)
	                             \sin \left( 2 \pi \omega_t \frac{\phi}{\phi_0} \right)
   .
 \label{eqn:e15}
\end{eqnarray}
The persistent current for single billiards exhibits a $\phi_0$ periodicity and changes from diamagnetic to paramagnetic by changing $k_F$.

In order to characterize the typical value of  the persistent current for single chaotic billiard, we define
\begin{equation}
   I^{typ}  \equiv  \left. \sqrt{ \left<  I^{(1)}(\phi)^2 \right> } \right|_{\phi/\phi_0=1/4}
   ,
 \label{eqn:e16}
\end{equation}
where the average $\left< \cdots \right>$ denotes an averaging over $k_F$ or some external parameter associated with the roughness of the boundary.
Upon averaging, the off-diagonal terms vanish due to the widely fluctuating phases, while the diagonal term survives the averaging.
Therefore we obtain
\begin{eqnarray}
   \left<  I^{(1)}(\phi)^2 \right> &=&
             \left( \frac{\sqrt{2} e v_F}{\pi } \right)^{2}
                 \sum_{t} \frac{R_T^2( L_t) \omega_t^2  }
                                        {L_t^2 \left| \det \left( M_t -I \right) \right|}
										\nonumber\\
             &\times&    \sin^2 \left( 2 \pi \omega_t \frac{\phi}{\phi_0} \right)
             .
 \label{eqn:e18}
\end{eqnarray}
To evaluate the sum of $t$, we order the orbits by their length $L_t$ and use the Hannay and Ozorio de Almeida (H-OdA) sum rule,~\cite{rf:H-OdA} i.e.,
\begin{equation}
   \sum_t \frac{1}{\left| \det \left( M_t -I \right) \right|} \cdots
   = 
   \int_0^{\infty} \frac{d L}{L} \sum_{\omega=1}^{\infty} P_L(\omega) \cdots
   .
 \label{eqn:e19}
\end{equation}
Here $P_L(\omega)$ denotes the classical distribution of $\omega$ that an orbit is of length $L$ and is approximately given by Gaussian distribution,~\cite{rf:Berry1,rf:Berry2} 
\begin{equation}
   P_L(\omega)=\frac{1}{\sqrt{2 \pi \overline{  \omega^2(L) }}} 
                         \exp \left( - \frac{\omega^2}{2 \overline{  \omega^2(L) }} \right)
                         ,
 \label{eqn:e20}
\end{equation}
with a variance increasing linearly with $L$, i.e.
\begin{equation}
  \overline{  \omega^2(L) } =\alpha \frac{L}{L_0}
  ,
 \label{eqn:e21}
\end{equation}
where $\alpha$ is a system dependent constant, and  $L_0$ is the length for the shortest periodic orbit.

In the following, we calculate $I^{typ}$ in high-temperature regimes, $T \geq T_{th}$,~\cite{rf:comment0} where  $T_{th}=\hbar v_F \pi / k_B L^*$  is the threshold temperature. ($L^*$ is the characteristic length for which periodic orbits can be taken as uniformly distributed in phase space.~\cite{rf:Agam}) 
In this regime, the thermal fluctuation washes out structures on the energy scale of the mean energy level spacing $\Delta$, and only short orbits contribute effectively to the persistent current.~\cite{rf:comment}
In this case one may exploit the asymptotic form $R_T(L) \approx 2(L/L_c) \exp (-L/L_c)$. Substituting eqs.~(\ref{eqn:e19}) and (\ref{eqn:e20}) into eq.~(\ref{eqn:e18}), and using the Poisson summation formula, lead to 
\begin{eqnarray}
  && \left<  I^{(1)}(\phi)^2 \right>
              =
             \left( \frac{2 e v_F}{\pi L_c} \right)^2 
              e^{-2/\omega_c}
              \left[
                       \frac{1} { \left( 1 - e^{-2/\omega_c} \right)^2 }
					   \right.
					   \nonumber\\
                     &+&
					 \left.
                        \frac{
                                  2 \left( 1 + e^{-4/\omega_c} \right)
                                  \sin^2 \left( 2 \pi  \frac{\phi}{\phi_0} \right)
                                  -
                                 \left( 1 - e^{-2/\omega_c} \right)^2 
                                 } 
                         { 
                           \left\{
                                     4 e^{-2/\omega_c}
                                     \sin^2 \left( 2 \pi  \frac{\phi}{\phi_0} \right)
                                     +
                                     \left( 1 - e^{-2/\omega_c} \right)^2 
                           \right\}^2
                         }
                         \right]
             ,
 \label{eqn:e22}
\end{eqnarray}
where $\omega_c=\sqrt{\alpha L_c/L_0}$.
Therefore, the typical value of persistent current is given by
\begin{equation}
I^{typ} =
                 \frac{2\sqrt{2}}{\pi} 
                 \frac{e v_F}{L_c} 
                 \frac{e^{-1/\omega_c} }
                 {
                    \left( 1 - e^{-2/\omega_c} \right)
                    \sqrt{ 1 + e^{-2/\omega_c} }
                 }
                 ; \quad T>T_{th}
   .
 \label{eqn:e23}
\end{equation}

The typical persistent current  [eq.~(\ref{eqn:e23})] scales as $I^{typ} \sim k_F \ell$, which is in agreement with the result of von Oppen and Riedel's dimensional analysis.~\cite{rf:Oppen}
In the case of integrable systems,~\cite{rf:Oppen,rf:RUJ1,rf:RUJ2} on the other hand, it scales as $I^{typ} \sim \left( k_F \ell \right)^{3/2}$.
This mean that the persistent current of integrable billiards is much larger than that of chaotic ones.

As in the case of diffusive~\cite{rf:disorder2,rf:disorder3,rf:disorder4,rf:disorder5} and integrable~\cite{rf:RUJ1} systems, $I^{(1)}$ gives a vanishing contribution to the grand canonical persistent current of an $ensemble$ of billiards with different sizes or electron fillings as soon as the dispersion in $k_F \ell$ is order of $2\pi$.
Therefore we need to go to the canonical term $\Delta F^{(2)}$ in the free-energy expansion.
Using eqs.~(\ref{eqn:e1}) and (\ref{eqn:e12}), for the canonical persistent current one finds
\begin{eqnarray}
  &&I^{(2)}(\phi) =
             -  \frac{e \hbar}{2 \pi m A} 
                 \sum_{s,s'} \frac{R_T( L_s) R_T( L_{s'}) }
                                        {   
                                          \sqrt{
										  \left| \det \left( M_s -I \right) \right|
                                          \left| \det \left( M_{s'} -I \right) \right|
										  }
                                        } \nonumber\\
                 &\times&
                 \left[
                      \left(
                              \omega_s + \omega_{s'}
                      \right)
                      \sin \left\{ 
                                       \frac{\bar{S}_s + \bar{S}_{s'} }{\hbar} 
                                       +
                                      2 \pi \left( \omega_s + \omega_{s'} \right) \frac{\phi}{\phi_0}
                              \right\} 
                   \right. \nonumber\\
                    &-&
                    \left.
                      \left(
                              \omega_s - \omega_{s'}
                      \right)
                      \sin \left\{ 
                                       \frac{\bar{S}_s - \bar{S}_{s'} }{\hbar} 
                                       +
                                      2 \pi \left( \omega_s - \omega_{s'} \right) \frac{\phi}{\phi_0}
                              \right\}
                 \right]
             ,
 \label{eqn:e25}
\end{eqnarray}
where $\bar{S}_s=S_s-\hbar \sigma_s \pi /2$.
For an ensemble with a large dispersion of sizes or $k_F$, only the Cooperon term ($s'$ is $s$ time reversed) survives the $k_F$ average, and we obtain
\begin{eqnarray}
   \left< I(\phi) \right> 
  & \approx&
   \left< I^{(2)}(\phi) \right> \nonumber\\
   &=&
   \frac{e \hbar}{\pi m A}
   \sum_{s} 
   \frac{ R_T^2( L_s) \omega_s }
           { \left| \det \left( M_s -I \right) \right|}
                 \sin \left( 4 \pi \omega_s \frac{\phi}{\phi_0} \right)
                 .
 \label{eqn:e26}
\end{eqnarray}
By using the H-OdA sum rule and the Gaussian winding number distribution, one obtains the average persistent current as
\begin{equation}
   \left< I(\phi) \right> 
   =
   \frac{\sqrt{2}e \hbar}{ \pi^{3/2} m A \omega_c}
   \sum_{n=1}^{\infty} 
                 f(\omega_c;n) n 
                 \sin \left( 4 \pi n \frac{\phi}{\phi_0} \right)
                ,
 \label{eqn:e27}
\end{equation}
where the function $f(\omega_c;n)$ is defined as
\begin{equation}
  f(\omega_c;n)
   =
   \int_0^{\infty} dx \ \frac{\sqrt{x}}{\sinh^2 x}
                                \exp \left( -\frac{n^2}{2 \omega_c^2  x}\right)
                 .
 \label{eqn:e28}
\end{equation}
Using the asymptotic expression $\sinh x\approx e^x/2$ valid in high-temperature regimes, $x=T/T_c>1$, this integral can be performed analytically.
The average persistent current for high temperature regime is thus given by
\begin{eqnarray}
\left< I(\phi) \right>  =
		 \frac{e \hbar }{\pi m A \omega_c} && 
		  \sum_{n=1}^{\infty} \left( 1+ \frac{2}{\omega_c} n \right) e^{-\frac{2}{\omega_c} n}
		  \nonumber\\
          &\times&
		  \sin \left( 4 \pi n \frac{\phi}{\phi_0} \right)
          ; \quad T>T_{th}
          .
 \label{eqn:e29}
\end{eqnarray}
$\left< I(\phi) \right> $ shows the halving of the flux period with respect to the nonaveraged persistent current $I(\phi)$, i.e., eq.~(\ref{eqn:e15}), characteristic of ensemble results.
Moreover the amplitude of $\left< I \right> $ for chaotic systems is positive, and scales as $\left< I \right> \sim \left( k_F \ell \right)^0$, which is much smaller than that of integrable systems,~\cite{rf:RUJ1} $\left< I \right> \sim k_F \ell$.
Note that $\left< I(\phi) \right> $ decreases rapidly with increasing temperature, and the higher harmonics die out more quickly.

In conclusion, we have studied the persistent current in quantum chaotic systems by using the Gutzwiller trace formula.
We derived semiclassical formulas of a typical persistent current $I^{typ}$ for single billiards and the averaged persistent current $\left< I \right>$ for an ensemble of billiards at finite temperature.
Moreover, we have shown that the persistent current amplitude scales as
$I^{typ} \sim k_F \ell$ and $\left< I \right> \sim \left( k_F \ell \right)^0$.
The former result coincides with von Oppen and Riedel's dimensional analysis.~\cite{rf:Oppen}
The different $k_F$ dependence of persistent current between integrable ($I^{typ} \sim \left( k_F \ell \right)^{3/2}$ and $\left< I \right> \sim k_F \ell $) and chaotic systems is attributed to the different contributions from nonisolated and from isolated periodic orbits.
Therefore, we can conclude that the quantum persistent current in ballistic systems becomes a tool to distinguish chaotic and regular classical dynamics experimentally. 
Unfortunately no experiment is yet available for chaotic billiards.~\cite{rf:exp}
We hope our results will encourage future experiment.

I would like to thank
B. Friedman, Y. Takane, and H. Tamura
for valuable discussions and comments.

\begin{references}
%
%
%
%
%
\bibitem{rf:PC}
For reviews of persistent current in mesoscopic systems, see 
H.F. Cheung, Y. Gefen and E.K. Riedel,
IBM J. Res. Dev. {\bf 32}, 359 (1988),
U. Eckern and P. Schwab,
Adv. Phys. {\bf 44}, 387 (1995)
and
Y. Imry, 
{\it Introduction to Mesoscopic Physics} 
(Oxford University Press, New York, 1997) Chap.~4.
%
%
%
%
\bibitem{rf:disorder1}
H.F. Cheung, E.K. Riedel and Y. Gefen,
Phys. Rev. Lett. {\bf 62}, (1989) 587.
%
%
%
%
\bibitem{rf:disorder2}
H. Bouchiat and G. Montambaux,
J. Phys. (Paris) {\bf 50}, (1989) 2695.
%
%
%
%
\bibitem{rf:disorder3}
F. von Oppen and E.K. Riedel,
Phys. Rev. Lett. {\bf 66}, (1991) 84.
%
%
%
%
\bibitem{rf:disorder4}
B.L. Altshuler, Y. Gefen and Y. Imry,
Phys. Rev. Lett.  {\bf 66}, (1991) 88.
%
%
%
%
\bibitem{rf:disorder5}
N. Argaman, Y. Imry and U. Smilansky,
Phys. Rev. B {\bf 47}, (1993) 4440.
%
%
%
%
\bibitem{rf:disorder6}
A. Szafer and B.L. Altshuler,
Phys. Rev. Lett.  {\bf 70}, (1993) 587.
%
%
%
%
\bibitem{rf:e-e1}
V. Ambegaokar and U. Eckern,
Phys. Rev. Lett. {\bf 65}, (1990) 381.
%
%
%
%
\bibitem{rf:e-e2}
A. Schmid,
Phys. Rev. Lett. {\bf 66}, (1991) 80.
%
%
%
%
\bibitem{rf:e-e3}
D. Ullmo, K. Richter, H.U. Baranger, F. von Oppen and R.A. Jalabert,
Physica E {\bf 1}, (1998) 268.
%
%
%
%
\bibitem{rf:QCmeso}
For a review of quantum chaos in mesoscopic systems, see 
{\it Chaos and Quantum Transport in Mesoscopic Cosmos}, 
edited by K. Nakamura,
Chaos Solitons Fractals 
{\bf 8} (1997) No.7/8.
%
%
%
%
\bibitem{rf:Oppen}
F. von Oppen and E.K. Riedel,
Phys. Rev. B {\bf 48}, (1993) 9170.
%
%
%
%
\bibitem{rf:RUJ1}
K. Richter, D. Ullmo and R.A. Jalabert,
Phys. Reps. {\bf 276}, (1996) 1.
%
%
%
%
\bibitem{rf:BerryTabor1}
M.V. Berry and M. Tabor,
Proc. R. Soc. Lond. A. {\bf 349}, (1976) 101.
%
%
%
%
\bibitem{rf:BerryTabor2}
M.V. Berry and M. Tabor,
J. Phys. A {\bf 10}, (1977) 371.
%
%
%
%
\bibitem{rf:Gutzwiller}
M.C. Gutzwiller,
in {\it Chaos and Quantum Physics}, edited by M.-J. Giannoni, A. Voros and 
J. Zinn-Justin (Elsevier, Amsterdam, 1991) p.~201.
%
%
%
%
%
\bibitem{rf:H-OdA}
J.H. Hannay and A.M. Ozorio de Almeida,
J. Phys. A {\bf 17}, (1984) 3429.
%
%
%
%
\bibitem{rf:Berry1}
M.V. Berry and M. Robnik,
J. Phys. A {\bf 19}, (1986) 649.
%
%
%
%
\bibitem{rf:Berry2}
M.V. Berry and J.P. Keating,
J. Phys. A {\bf 27}, (1994) 6167.
%
%
%
%
\bibitem{rf:comment0}
At $low$ temperature  $T \ll \Delta/k_B$, where $\Delta \ll k_BT_{th}$, longer periodic orbits become more important and make a large contribution to the persistent current.
In this limit the problem becomes nonperturbative, so that one must take into account off-diagonal terms between very long classical trajectories.
Moreover in this limit the thermodynamical expansion [eq. (8)] breaks down.
%
%
%
%
\bibitem{rf:Agam}
O. Agam,
J. Phys. I France {\bf 4}, (1994) 697.
%
%
%
%
\bibitem{rf:comment}
In this paper discussion will be confined to chaotic billiards for which the boundary is complicated enough (see Fig.~1) so that one may consider the corresponding set of the $shortest$ periodic orbits as covering the phase space.
Therefore we can use the H-OdA sum rule and the Gaussian winding number distribution even for the shortest periodic orbits.
%
%
%
%
\bibitem{rf:RUJ2}
R.A. Jalabert, K. Richter and D. Ullmo,
Surf. Sci. {\bf 361}/{\bf 362}, (1996) 700.
%
%
%
%
\bibitem{rf:exp}
For an experimental study of the persistent current for $integrable$ systems, see
D. Mailly, C. Chapelier and A. Benoit,
Phys. Rev. Lett.  {\bf 70}, (1993) 2020.
%
%
%
%
%
%
%
%
\end{references}
\end{document}